\documentclass{elsart}
\usepackage{amssymb,epsfig}
\begin{document}
\begin{frontmatter}

\title{Low Temperature Neutron Diffraction Study of MnTe}
\author[gu]{J. B. C. Efrem D'Sa}, \author[gu]{P. A. Bhobe}, \author[gu]{K. R. Priolkar\corauthref{krp}}\ead{krp@unigoa.ac.in},
\author[barc1]{A. Das}, \author[iaea]{S. K. Paranjpe\thanksref{barc}}, 
\author[gu]{R. B. Prabhu} and \author[gu]{P. R. Sarode}
\corauth[krp]{Corresponding author}
\address[gu]{Department of Physics, Goa University, Taleigao-Plateau, Goa, India 403 206}
\address[barc1]{Solid State Physics Division, Bhabha Atomic Research Centre, Trombay, Mumbai, India 400 085}
\address[iaea]{Physics Section, NAPC, International Atomic Energy Agency, Wagramer Strasse 5, A-1400 Vienna, Austria}
\thanks[barc]{Formerly at Solid State Physics Division, Bhabha Atomic Research Centre, Trombay, Mumbai, India 400 085}

\begin{abstract}
Investigation of transport and magnetic properties of MnTe at low temperatures showed anomalies like negative coefficient of resistance below 100K and a sharp rise in susceptibility at around 83K similar to a ferromagnetic transition. Low temperature powder neutron diffraction experiments were therefore carried out to understand the underlying phenomena responsible for such anomalous behavior. Our study indicates that the rise in susceptibility at low temperatures is due to strengthening of ferromagnetic interaction within the plane over the inter plane antiferromagnetic interactions.  
\end{abstract}


\end{frontmatter}

\section{Introduction}
Compounds of 3d transition metals with VIB elements have a good combination of both electrical and magnetic properties. Because of their interrelation and incomplete d-shell of the component transition metals, they exhibit a variety of structural, magnetic, electrical, optical and thermal properties. As a consequence of this diversity they have excited considerable interest. They generally crystallize in the hexagonal NiAs type structure although a few of them possess cubic NaCl type structure. These NiAs type structures are important because of the interesting magnetic as well as electrical properties arising from the incomplete d shells of the component transition metals \cite{jek} and are of great interest both from experimental and theoretical point of view \cite{kuzm,barn}. Besides transformation between magnetic phases, in some cases the transitions accompany a change in crystal structure. Special attention is devoted to the anomalies of magnetic, elastic and electrical properties of these substances which arise at the phase transition and the nature of which has not yet been understood. In particular all these anomalies have been observed in compounds containing manganese viz., MnAs, MnSb, MnTe etc. \cite{sand}. Manganese chalcogenides, MnS, MnSe and MnTe are antiferromagnetic compounds with transition temperatures increasing with molecular weight \cite{bize,squi}. In the case of MnS and MnSe the stable crystal structure is cubic NaCl type and these are wide gap insulators. \cite{alle,deck,oguc}. On the other hand MnTe crystallizes in the hexagonal NiAs type structure \cite{wyck} and it is one of the few semiconductors among the 3d transition metal compounds. MnTe has electrical and magnetic properties similar to those of NaCl type chalcogenide magnetic insulators but has a crystal structure which normally supports metallic conductivity. It is a p-type semiconductor with a direct band gap  of about 1.3 eV. It orders antiferromagnetically at N$\acute{e}$el temperature, T$_{N}\sim$ 310K \cite{uchi,yada,bane,kuni}. Changes in the temperature dependence of resistivity near T$_{N}$ have been attributed to spin disorder scattering i.e. scattering by magnons \cite{genn}, a behaviour normally seen in ferromagnetic materials \cite{hass}. The anomalies near the T$_N$ in the thermoelectric power \cite{wass} and the Hall effect \cite{mara} are all related to the interaction of the charge carriers with the localized spins i.e. magnon drag effect \cite{genn}, a phenomenon similar to phonon drag in non magnetic semiconductors. MnTe behaves like an ionic crystal since the tellurium states prevail in the valence band, while the manganese states prevail in the conduction band. On the other hand dispersion of the bands indicate strong covalent effects. Thus MnTe is a crossroad material sharing properties of both broad band semiconductors and ionic salts \cite{alle}. 

In this paper we report the results of the study of crystal and magnetic structure of MnTe at low temperatures. The transport and magnetic measurements such as resistivity and D.C. susceptibility carried out on polycrystalline MnTe showed anomalous behaviour at low temperatures. A rise in resistivity below 100K accompanied by a rise in D.C. susceptibility at around the same temperature ($\sim$ 83K), indicative of a ferromagnetic like transition was observed. Neutron diffraction studies in the temperature range 10K to 300K were carried out to understand these anomalies and to elucidate the low temperature crystal and magnetic structure of this compound.

\section{Experimental}
Polycrystalline samples of MnTe were prepared by the standard solid state route. The stoichiometric amount of the starting materials, Mn and Te in metallic form were ground, pelletized and vacuum sealed in an quartz ampoule below 10$^{-6}$ Torr which was then annealed at a temperature of 750$^\circ$C for 10 days. The sample formed was single phase and homogeneous with NiAs structure as confirmed by X-ray diffraction. The lattice parameter values calculated from the X-ray diffraction pattern were a = 4.190$\pm$0.001  \AA~~ and c = 6.751$\pm$0.001 \AA. 

 The sample was then characterized by temperature dependent resistivity measurement using the standard four probe technique. The DC susceptibility measurement was performed on the Faraday balance. In order to understand the nature of the anomalies seen in resistivity and susceptibility of MnTe neutron diffraction study has been performed on this sample. Neutron diffraction experiments were carried out in the temperature range 10K to 300K at the wavelength of 1.242 \AA~~ using the Profile Analysis Diffractometer at Dhruva reactor, Trombay.  

\section{Results and Discussion}
\begin{figure}[h]
\epsfig{file=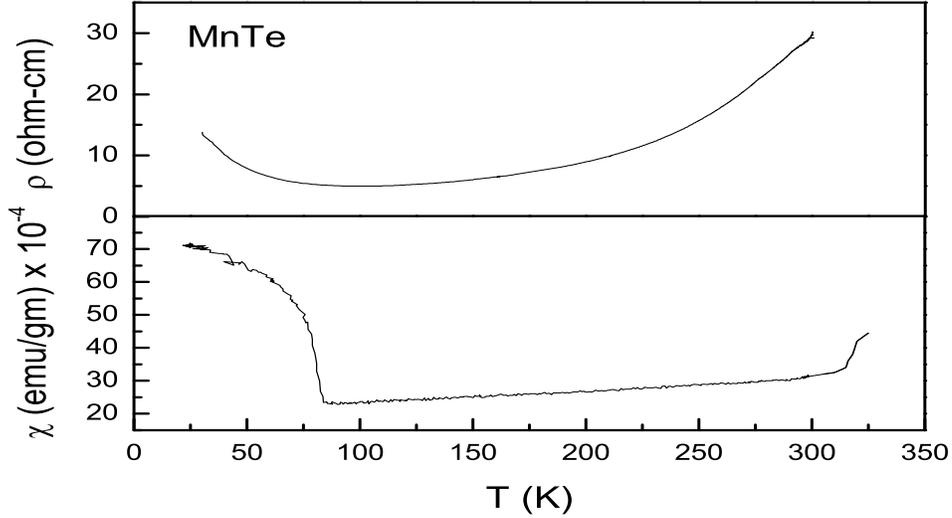, width=14cm, height=8cm}
\caption{Resistivity and Susceptibility of MnTe as a function of temperature}
\label{fig1}
\end{figure}

The resistivity and D. C. susceptibility as a function of temperature for MnTe is shown in figure \ref{fig1}. MnTe is a semiconductor at room temperature and the resistivity is expected to increase with decreasing temperature. However, the positive coefficient of resistivity seen below the magnetic ordering temperature T$_{N}\sim$ 310K is due to the spin disorder scattering with a large contribution from the influence of magnon drag \cite{hass}. As the temperature is further lowered, a rise in resistivity is observed below 100K. The D.C. susceptibility exhibits a sudden drop at $\sim$ 310K indicating an antiferromagnetic order in agreement to the earlier reports. The anomaly observed here is the sharp rise in susceptibility around 83K suggestive of a ferromagnetic like order.

\begin{figure}[h]
\centering
\epsfig{file=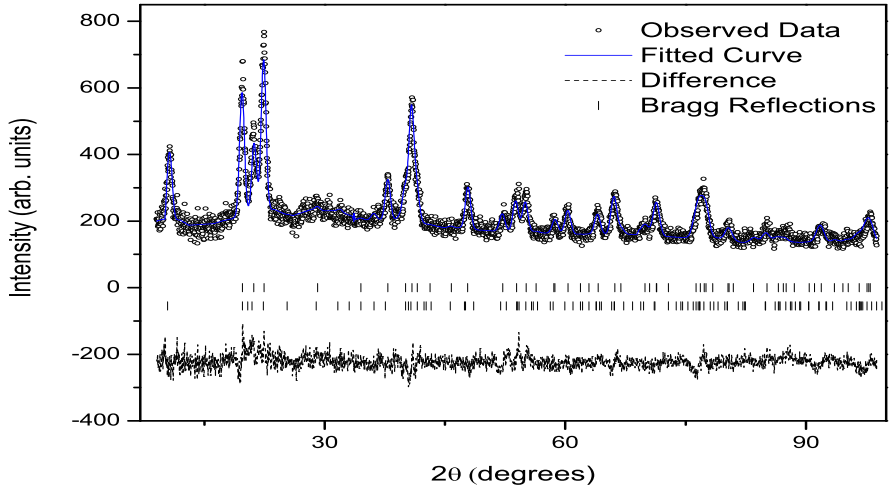, width=14cm, height=8cm}
\caption{Neutron Diffraction Pattern of MnTe at 300K. The top set of vertical lines refer to nuclear Bragg reflections whereas the bottom set indicate positions of magnetic Bragg reflections.}
\label{pattern}
\end{figure}

\begin{figure}[h]
  \centering
  \epsfig{file=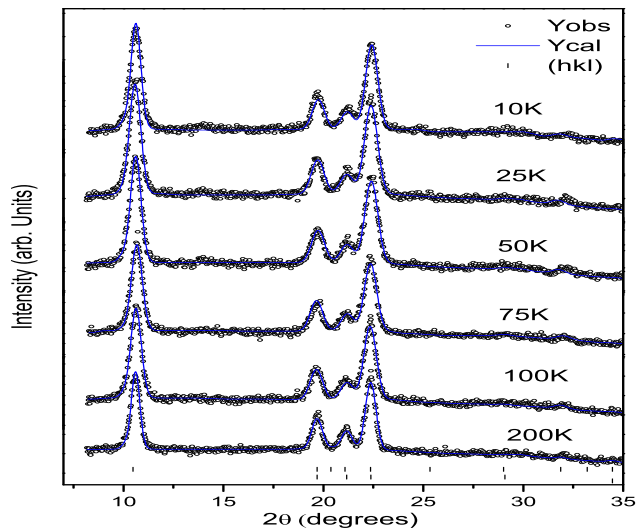, width=10cm, height=8cm}
  \caption{Neutron Diffraction pattern of MnTe at various temperatures.The top set of vertical lines refer to nuclear Bragg reflections whereas the bottom set indicate positions of magnetic Bragg reflections.}
  \label{alltemp}
\end{figure}

In order to understand the origin of these anomalous behaviour at low temperature, powder neutron diffraction measurements on MnTe as a function of temperature were carried out. Profile refinement of the mentioned diffraction data at various temperatures was carried out using the FULLPROF program \cite{fp}. The 300K pattern recorded for the entire setting of the detector i.e. from 2$\theta$ = $8^\circ$ to $96^\circ$ is shown in figure \ref{pattern}. An intense Bragg peak due to the antiferromagnetic ordering of Mn sub-lattice is seen at 2$\theta \approx  10^\circ$ in the 300K pattern along with the peaks corresponding to the NiAs type chemical structure. This pattern shows MnTe to be hexagonal with lattice parameters a = 4.193$\pm$0.001\AA,~ c = 6.752$\pm$0.001\AA~, in close agreement to those obtained from X-ray diffraction pattern [not shown]. The structural parameters obtained from Rietveld refinement of the room temperature pattern are presented in Table \ref{riet1}. Since the magnetic form factor for neutrons falls drastically at higher angles, the low temperature diffraction data are limited upto 2$\theta$ = $35^\circ$ and are presented in the figure \ref{alltemp}. The patterns could be fitted well with two phase refinement of nuclear and magnetic phases of MnTe. The magnetic cell of MnTe could be viewed as Mn moments aligned ferromagnetically in the basal plane and these planes stacked antiferromagnetically along the c - axis which is in agreement with those reported in literature \cite{kuni,gree}. The magnetic moment per Mn ion is found to be 0.92$\mu_{B}$ at 300K. The parameters obtained at room temperature were used as inputs to refine the neutron diffraction data at lower temperatures. The parameters refined were cell parameters and magnetic moment parameters, M$_x$ and M$_y$. The fitted patterns to the experimental data at different temperatures are shown in figure \ref{alltemp} as solid lines. Absence of any extra peaks in the low temperature neutron diffraction patterns clearly rule out the possibility of any structural transition.   

\begin{table}[h]                               
\caption{Position coordinates, Isotropic temperature factor (B$_{iso}$), Occupancy/formula unit (Occ./f.u.) and R-factors of Rietveld refinement of 300K neutron diffaction pattern of MnTe.} 
\label{riet1}  
\begin{center}
\begin{tabular}{|cccccc|} 
\multicolumn{6}{l}{Space Group: P6$_3$/mmc}\\
\hline  
Atom & $x$ & $y$ & $z$ &B$_{iso}$(\AA$^2)$ & Occ./f.u. \\  
\hline
Mn & 0 & 0 & 0 & 0.84(1) & 1.00(1)\\  
Te & 1/3 & 2/3 & 1/4 & 0.57(1) & 0.98(1)\\
\hline  
\multicolumn{1}{l}{R$_p$ = 5.8} & \multicolumn{2}{l}{R$_{wp}$ = 7.33} & \multicolumn{2}{l}{R$_{exp}$ = 6.48}\\
\multicolumn{2}{l}{Bragg R = 14.7} & \multicolumn{4}{l}{Magnetic R = 10.8} \\
\hline
\end{tabular}  
\end{center}

\vspace{0.2cm}
{\small The number in the bracket represents uncertainty in the last digit.}
\end{table}  

\begin{figure}[h]
  \centering
   \epsfig{file=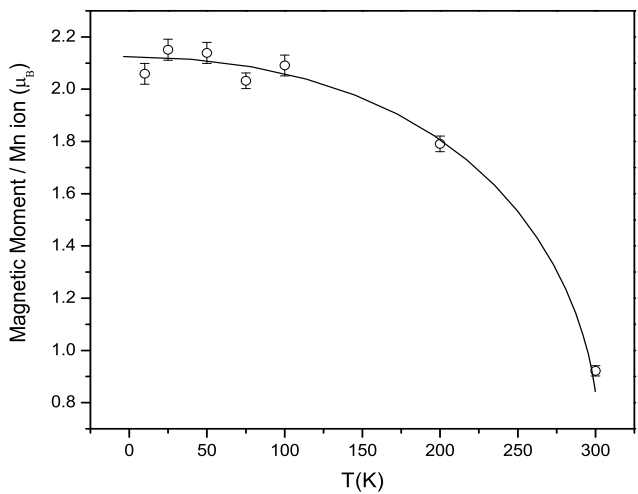, width=10cm, height=8cm}
  \caption{Variation of Magnetic Moment with temperature. Solid line is just a line through the points}
  \label{magmom}
\end{figure}

A plot of magnetic moment per Mn ion estimated from our analysis is shown in figure \ref{magmom}. It is seen that the magnetic moment decreases from 2.06$\mu_{B}$ at 10K to 0.92$\mu_{B}$ at 300K. The variation of magnetic moment with temperature shows the Brillouin function type of dependence. The expected value of the ordered moment at T = 0 K is 5$\mu_B$. Instead, a reduced moment of $\sim$ 2$\mu_B$ is observed. This is perhaps due to large dispersion of Mn bands in MnTe which results in Mn 3d band being more than half-filled. The rise in susceptibility below 83K could be due to a tilt in magnetic moment away from the basal plane. However, a tilt in magentic moment vector away from the basal plane would have resulted in increase in intesity of  magnetic $\{102\}$ reflection (at 2$\theta$ = 22.35$^\circ$) and decrease in the intensity of magnetic $\{002\}$ reflection (at 2$\theta$ = 10.46$^\circ$). In the present case, the intensity of magnetic $\{102\}$ reflection as a function of temperature, exactly follows the intensity of magnetic $\{002\}$ reflection. This leads us to believe that there is no tilt in the magnetic moment vector away from the basal plane. Therefore other possibilities have to be considered to explain the increase of susceptibility below 83K. 
 
\begin{figure}[h]
  \centering
   \epsfig{file=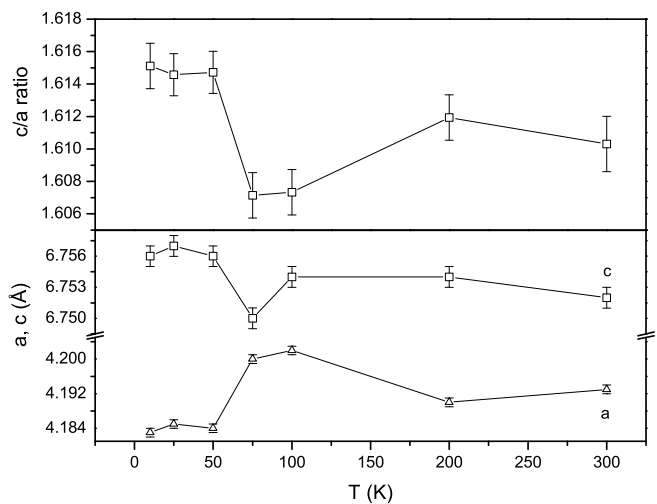, width=10cm, height=8cm}
  \caption{Variation of lattice parameters and the c/a ratio with temperature.}
  \label{cell}
\end{figure}

Owing to the magnetic stucture of MnTe, the magnetic interactions seem to be strongly associated with the structural parameters and any change in the latter would affect the overall magnetic interactions. In the present study, as can be seen in figure \ref{cell}, $a$ and $c$ and therefore the $c/a$ ratio remain constant upto around 200K. The $c/a$ then decreases upto 75K with $c$ and $a$ having a minimum and maximum values respectively. Below 75K $c/a$ ratio increases sharply with decreasing temperature due to an increase in $c$ and decrease in $a$ values. The Mn-Mn interatomic distance in the basal plane is equal to $a$ while along the perpendicular axis it is given by $c/2$. It can be clearly seen from figure \ref{cell} the change in $c$ below 75K is much smaller than the corresponding change in $a$. Therefore the decrease in $a$ would strengthen the ferromagnetic interactions between the Mn ions in the basal plane at the same time the increase in $c$ weakens the antiferromagnetic interactions. Such a weakening of inter planar coupling can also explain the small rise seen in resistivity at low temperatures. It may be noted that no ferromagnetic signature is detected in neutron diffraction patterns below 75 K that could be attributed to any structural transition. Hence the observed anomalies below 100K in the resistivity and susceptibility measurements are due to a magneto-elastic coupling wherein the Mn ions are slightly  displaced without producing any drastic change in the overall crystal structure.   

\section{Conclusion}
The resistivity and susceptibility measurements as a function of temperature for MnTe were carried out. The upturn in the susceptibility at around 83K hinted towards some kind of structural transition. However, a detailed analysis of the powder neutron diffraction measurements in the temperature range 10K to 300K do not show any evidence of an extra phase to be present within the limited statistics of the data. The rise in susceptibility below 83K could be explained to be due to a magneto-elastic coupling which strengthens intra planar ferromagnetic interactions relative to inter planar antiferromagnetic interactions.

\section*{Acknowledgments}
This work is supported by Inter University Consortium for DAE facilities (IUC-DAEF) under the project CRS-M-63.

\end{document}